\pdfoutput=1

\documentclass{iopart}
\usepackage{iopams}
\usepackage[english]{babel}
\usepackage{cite} 
\usepackage{textcomp}
\usepackage{graphics}
\usepackage{graphicx}
\usepackage{anysize}
\usepackage{amssymb}
\usepackage{subfigure}
\usepackage{float}
\usepackage[normalem]{ulem}

\usepackage{array}

\newcommand{\text}[1]{{\rm #1}}

\usepackage{color}
\definecolor{darkgreen}{rgb}{0.0, 0.4, 0.0}
\definecolor{oooo}{rgb}{0.99, 0.5, 0.01}

\begin{document}
\title{Flow in an hourglass: particle friction and stiffness matter}

\author{
Tivadar Pong\'o,\textit{$^{1,2}$}
Vikt\'oria Stiga,\textit{$^{1}$}
J\'anos T\"or\"ok,\textit{$^{3,4}$}
S\'ara L\'evay,\textit{$^{4}$}
Bal\'azs Szab\'o,\textit{$^{1}$}
Ralf Stannarius,\textit{$^{5}$}
Ra\'ul Cruz Hidalgo,\textit{$^{2}$} and
Tam\'as B\"orzs\"onyi$^{\ast}$\textit{$^{1}$}
}

\address{
$^1$Institute for Solid State Physics and Optics, Wigner Research Centre for
Physics, P.O. Box 49, H-1525 Budapest,\\
$^2$F\'isica y Matem\'atica Aplicada, Facultad de Ciencias, Universidad de Navarra, Pamplona, Spain\\
$^3$MTA-BME Morphodynamics Research Group, Budapest University of
Technology and Economics, Budapest, Hungary\\
$^4$Department of Theoretical Physics, Budapest University of
Technology and Economics, Budapest, Hungary\\
$^5$Institute of Physics, Otto von Guericke University, Magdeburg, Germany
  }

\ead{borzsonyi.tamas@wigner.hu}


\begin{abstract}
Granular flow out of a silo is studied experimentally and numerically.
The time evolution of the discharge rate as well as the normal force (apparent weight)
at the bottom of the container is monitored.
We show, that particle stiffness has a strong effect on the qualitative features of silo
discharge. For deformable grains with a Young's modulus of about $Y_m\approx 40$ kPa
in a silo with basal pressure of the order of 4 kPa lowering the friction coefficient leads to 
a gradual change in the discharge curve:
the flow rate becomes filling height dependent, it decreases during the discharge process.
For hard grains with a Young's modulus of about $Y_m\approx 500$ MPa the flow rate
is much less sensitive to the value of the friction coefficient.
Using DEM data combined with a coarse-graining methodology allows us to compute
all the relevant macroscopic fields, namely, linear momentum, density and stress tensors.
The observed difference in the discharge in the low friction limit is connected to a strong 
difference in the pressure field: while for hard grains Janssen-screening is effective, leading to 
high vertical stress near the silo wall and small pressure above the orifice region, for deformable 
grains the pressure above the orifice is larger and gradually decreases during the discharge process.
We have analyzed the momentum balance in the region of the orifice (near the location of 
the outlet) for the case of soft particles with low friction coefficient, and proposed a 
phenomenological formulation that predicts the linear decrease of the flow rate with decreasing 
filling height.

\end{abstract}


\vspace*{40mm}
\noindent
Published in:
\medskip

\noindent
New J. Phys. \textbf{23} 023001 (2021)\\
\texttt{https://doi.org/10.1088/1367-2630/abddf5} \qquad (open access)
\newpage

\section{Introduction}

Gravity driven discharge of a granular material from a silo or hopper is a frequent operation
in various industrial procedures
\cite{nedderman1982,jaeger1992,jaegerRMP1996,jaegerPT1996,kakalios2005,janda2012,tighe2007,campbell2006}.
Materials involving deformable and/or viscoelastic grains with low surface friction pose 
new challenges for controlling the flow of such samples.
Several experimental and numerical studies have shown that for frictional hard grains the
discharge rate of a silo is constant and independent of the filling height
\cite{nedderman1982,mankocGM2007,martinez2008,ahnPT2008,jin2010,baleviciusPT2011,oldal2012},
if the orifice diameter $D$ is larger than about 5 particle diameters \cite{zuriguel2005}.
This feature is very useful in engineering applications and lead to the emergence 
of simple devices such as the hourglass.
More precisely, the flow rate was found to be constant until near the end of the discharge
when the shape of the bed surface (often having a form of a funnel) starts perturbing the outflow.
These results hold for such particles as 
spherical beads of various materials (glass, steel, lead, plastic) 
\cite{mankocGM2007,martinez2008,ahnPT2008,perge2012,koivisto2017,koivistoSM2017},
nearly spherical particles (e.g.~peas) \cite{baleviciusPT2011},
grains with irregular shape (e.g.~sand) \cite{tighe2007}, or slightly anisometric grains
(wheat, corn, soybeans, poppy seeds, oat) \cite{jin2010,oldal2012}.
The fact that the flow rate was found to be constant for a wide range of materials facilitated
the characterization of the outflow as a function of the orifice diameter $D$.
The flow rate $Q$ in a three-dimensional (3D) silo was found to vary approximately as $D^{5/2}$, 
where the best fit was obtained either using  an effective orifice diameter in place of $D$,
taking into account the grain size $d$ \cite{tighe2007,beverloo1961,nedderman1982},
or taking into account the slight density decrease near the orifice with decreasing orifice size
\cite{mankocGM2007}.
This so-called Beverloo scaling was experimentally tested in various systems and was shown to work
even at increased or decreased gravitational acceleration \cite{dorbolo2013}.

It is also known that the weight of the material in the silo is partly supported by the frictional 
contacts with the walls. As a consequence, the basal pressure $p_\text{b}$ (i.e. the normal stress 
measured at the bottom of the silo) saturates with increasing fill height $h$.
This phenomenon was first systematically investigated by Janssen \cite{janssen1895,sperl2006},
who confirmed the earlier measurements of Huber-Burnand \cite{huber1929},
Hagen \cite{hagen1852,tighe2007}, and Roberts \cite{roberts1884}. The so-called Janssen
effect is also present when the contacts are mobilized (experiments with moving walls)
\cite{vanel1999,bertho2003} and the $p_\text{b}(h)$ curve is best described by a combination
of a linear section (hydrostatic behaviour) at the bottom of the silo and an exponential
saturation in the upper section \cite{vanel1999,vanel2000}, where the saturation length
is comparable to the silo diameter $D_\text{silo}$.

The height-independent constant basal pressure would serve as a good argument for the constant 
flow rate as several authors mentioned
\cite{jaeger1992,jaegerRMP1996,kakalios2005,campbell2006,mankocGM2007,hilton2011}.
For comparison, in a cylinder filled with a liquid the discharge rate is clearly determined
by the pressure (compare, for example, the clepsydra), which is linearly increasing with the 
height of the liquid column. In the low viscosity limit, the conservation of energy leads to 
a flow rate $Q$ proportional to $\sqrt{h}$ (Torricelli's law).
For large viscosities, dissipation dominates near the outlet, leading to $Q \propto h$.
Investigations on granular flows showed, however, that the discharge rate is independent not
only on $h$, but also on the silo diameter $D_\text{silo}$ in case $D_\text{silo}>2.5D$ and
$D_\text{silo}>D+30d$ \cite{brown1960,franklin1955}.
Thus, a simple relation between $Q$ and $p_\text{b}$ has to be dropped, since according to the above
described Janssen effect, $p_\text{b}$ clearly increases with $D_\text{silo}$. 
Pacheco-Martinez et al.~also showed \cite{martinez2008} that slightly vibrating the hopper wall 
one can get hydrostatic conditions in the pressure profile (i.e.~Janssen screening disappears), but 
when the orifice is opened, Janssen screening reappears, and the silo still discharges the same
way as without vibration. 
Other recent experiments showed that different filling protocols might lead to different
pressure profiles, but the flow rate was found to be independent of the basal pressure
\cite{perge2012,peralta2017}.

The mechanism leading to a constant discharge rate of a granular material is still the subject
of active research. Detailed experimental and numerical investigations are carried out to
analyze the flow near the orifice, where the nature of dissipation clearly changes.
There is a region with paraboloid or hemispherical shape below which dissipation
strongly drops. This was first identified by Hagen \cite{hagen1852}.  Based on this, 
Brown and Richards introduced the concept of a free-falling arch signifying the
region where energy dissipation decreases to a minimum \cite{brown1960}. At this surface,
the stress in flow direction vanishes and from here the grains are freely falling
\cite{nedderman1992}.
Similar reasoning lead to the development of an alternative approach called the hourglass
theory, which predicts the discharge rate quantitatively \cite{savage1967,davidson1973}.
Recent works tested and refined these ideas
\cite{nedderman1982,hilton2011,oldal2012,mankocGM2007,rubiolargo2015,darias2020}.
Among these, Darias et al.~derived differential equations based on energy balance and
the so called $\mu(I)$ rheology \cite{darias2020}. Their results are compatible with the
Beverloo law, and the authors found an increase in the flow rate when the interparticle friction
is reduced. They also validated their theory numerically by 3D Discrete Element Model (DEM)
simulations.

Other recent investigations tested the limitations, i.e.~identified conditions where the flow
rate is not constant anymore. Introducing an interstitial fluid clearly changes the dynamics,
resulting in an increased flow rate near the end of the discharge process
\cite{wilson2014,koivistoSM2017}.
Recently, this effect was shown to be present -- although with a much smaller amplitude -- in
dry granular flows of glass beads in case the particles were sufficiently small ($d<1$ mm)
\cite{koivisto2017}.
For $d=2$ mm glass beads, the flow rate was again found to be constant \cite{koivistoSM2017}.
Another investigation pointed out that applying a large stationary external force leads to
an increased flow rate during the discharge of a dry granular material \cite{madridEPL2018}.

Some of the earlier numerical investigations mentioned that decreasing the frictional
damping leads to height dependent (i.e.~not constant) flow rates
\cite{hirshfeld1997,hirshfeld2001}.
Another numerical work reported strongly changing pressure conditions, but only slightly
increasing flow rates with decreasing wall friction \cite{jofriet1997}.
Using discrete element simulations, Balevicius indicated decreasing flow rates with increasing
interparticle friction in a quasi-2D system with limited size, and a time dependent flow
rate \cite{balevicius2007}.
In the above-mentioned recent work by Darias et al.~\cite{darias2020} on a 3D system, a
time dependent (slightly decreasing) flow rate is noticeable for their lowest value ($\mu=0.1$)
of the friction coefficient.
Langston et al.~\cite{langston1995} investigated the case of frictional grains ($\mu=0.6$) and 
found that the discharge rate was rather insensitive to the stiffness of the interparticle interactions. 
In a different numerical approach, Staron et al.~implemented a plastic rheology in a
2D Navier-Stokes solver (following the so called $\mu(I)$ rheology or constant friction)
\cite{staron2012}, which showed a transition from granular-like (constant flow rate) to
liquid-like (decreasing flow rate) behaviour with decreasing friction below about $\mu=0.3$.

In the present work, we show that the discharge behaviour of low friction soft hydrogel
beads strongly differs from the case of hard frictional grains described by a constant
flow rate. We quantify the difference in laboratory experiments by measuring the flow rate
and the normal force exerted on the bottom of the silo during the discharge process for 
both low friction soft and frictional hard grains. We perform numerical (Discrete Element Method) 
simulations to explore the effect of grain softness and surface friction 
on the flow rate as well as on the stress conditions inside the silo.
Our work was motivated by 
recent investigations on clogging statistics of soft particles in two-dimensional hoppers
\cite{ashourPRF2017,hong2017,harth2020} and X-ray tomographic measurements on the flow field 
of soft particles in a three-dimensional hopper \cite{stannarius2019,stannarius2019-2}, all of 
which noticed a fill height dependence of the clogging probability or the flow properties.

\section{Experimental system}
\label{exp}

\begin{figure}[h!]
\begin{center}
  \includegraphics[width=12cm]{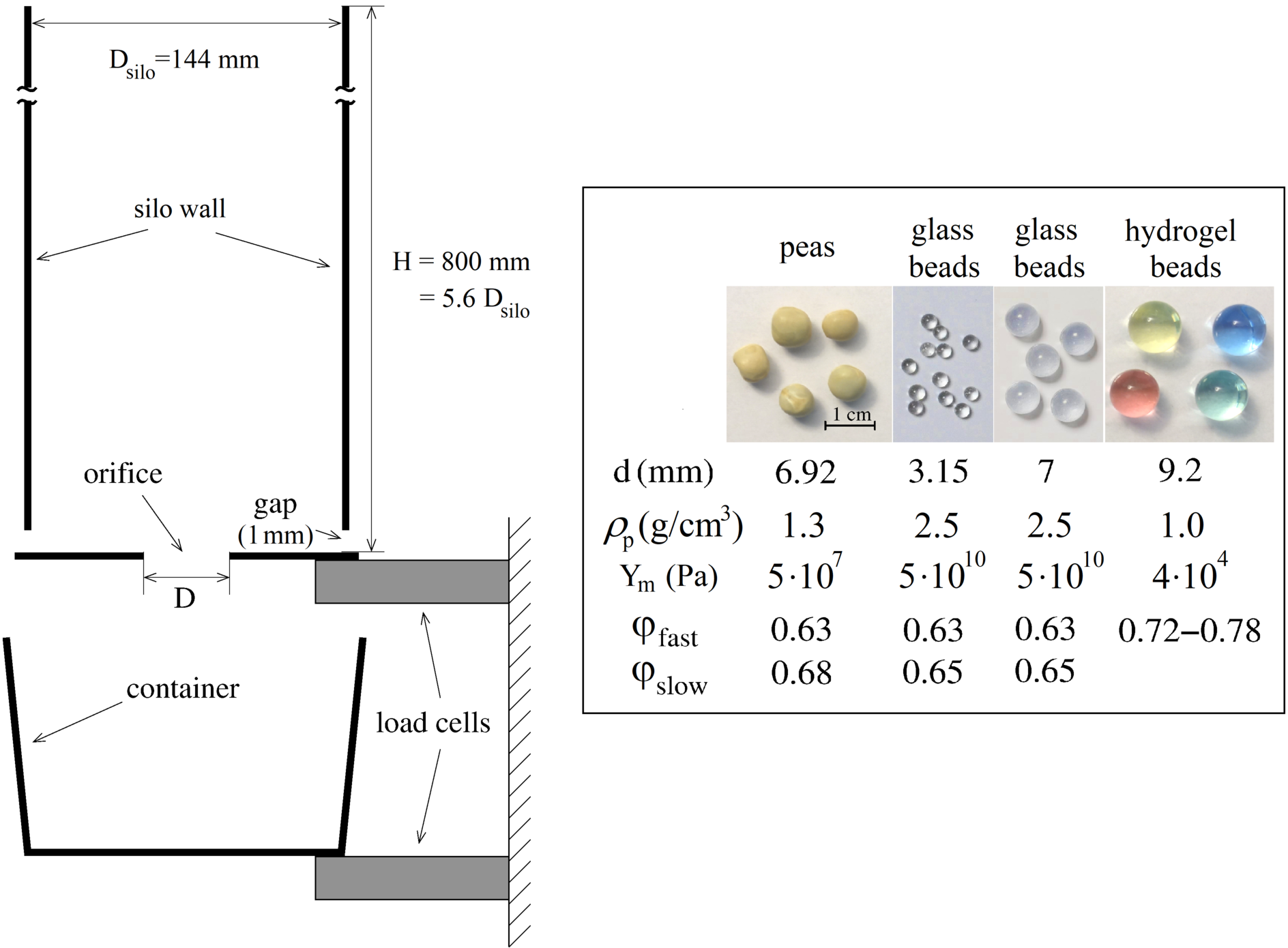}  
\end{center}
\vspace{-0.5cm}
  \caption{Sketch of the experimental setup and photographs of the samples, with the
          grain diameter $d$, particle density $\rho_p$, and Young modulus $Y_m$ indicated below.
      The initial packing fractions in the silo corresponding to fast and slow filling,
      $\varphi_\text{fast}$ and $\varphi_\text{slow}$ are also given.
  }
  \label{setup}
\end{figure}

An acrylic cylindrical silo with inner diameter $D_\text{silo}=144$ mm and height
$H=800$ mm was used. The bottom plate was physically disconnected from the silo with
a gap of approximately 1 mm, and it was held by a load cell enabling a continuous
monitoring of the force $F_\text{b}$ exerted on the bottom (see Fig.~\ref{setup}). 
Two types of filling protocols were used: pouring the grains into the silo with faster
($500\ \rm{cm}^3/\rm{s}$) and slower ($50\ \rm{cm}^3/\rm{s}$) speed. Lower pouring rates
resulted in slightly (a few percent) denser packing of the granular bed (see Fig.~\ref{setup} 
for the data). The silo was discharged through a circular orifice of diameter $D$ in the 
middle of the bottom plate. The typical discharge time was in between 50-100 s.
During discharge, both the force $F_\text{b}$ exerted on the bottom of the silo and the 
discharged mass were continuously measured by the load cells at a sampling rate of 1 kHz. 
The data series were then smoothed by a convolution with a Bartlett (triangle) window of 
a width of 1 s to reduce noise. The data for the discharged mass were differentiated and smoothed 
again with the same smoothing procedure to obtain the flow rate $Q$. The value of $Q$ is presented 
in grains/s. For this purpose, the average unit mass of the individual grains was determined by 
measuring the mass of 200 grains. The evolution of the fill height $h$ during discharge 
of hard grains was calculated from the discharged mass data assuming a linear relation 
between the height and the mass in the silo. For soft grains, pressure induces compression
of the material, so the discharged mass--fill height relation was approximated by a third order 
polynomial which was previously calibrated.

Four granular samples have been used in the experiments: nearly spherical pea seeds 
(supplier: Gyari-Mag BT, Hungary),
two samples of spherical glass beads (with diameters  
in the range of $2.85$ mm $< d < 3.45$ mm and $ 6.7$ mm $ < d < 7.3$ mm,
both from Sigmund-Lindner GMBH, Germany), and spherical hydrogel beads (Happy Store, Nanjing, China) 
(see Fig.~\ref{setup}).
For the glass bead samples, we used two versions: clean and lubricated. Lubrication was 
achieved by spraying  the glass beads ($\approx$ 20 kg) before pouring them into the silo 
as well as the inner surface of the silo with about 500 ml of silicone oil (Motip).
The table also indicates the packing fractions resulting after fast or slow
filling of the silo. For the hydrogel beads, the packing fraction depends on the magnitude
of the contact forces, thus it changes with height \cite{stannarius2019}.

Three samples (peas, clean glass beads with $d=3.15$ mm and $d=7$ mm) are considered
frictional hard grains with a friction coefficient of about 0.3-0.5, while the hydrogel 
balls are soft and have very small interparticle friction coefficient.
By measuring the Hertzian contact diameters, the Young's modulus $Y_m$ of our hydrogel
spheres was found to be between 30 and 50 kPa \cite{ashourPRF2017},
with a slightly softer outer part than the core.
This is similar to the results of other measurements with hydrogel beads, which yielded
a Young's modulus of $Y_m\approx 20$ kPa \cite{dijksmanRSI2017}).
In earlier studies, the interparticle friction coefficient of hydrogel beads was found to be
$\mu_p<0.03$ \cite{broduNC2015}.
We will show below that the discharge of the low friction soft hydrogel particles
is very different from the frictional hard grains. This motivated us to reduce
the surface friction of our glass bead samples by spraying
silicone oil on the surface of the beads as well as on the silo wall.

The friction coefficient for hard surfaces with a lubricating layer depends on the 
normal force and the sliding velocity \cite{persson1998}, so we carried out experiments 
on inclined planes to measure this friction coefficient.
In the {\it first set} of experiments, a wooden block of a mass of 188 g was used with three glass
beads glued on it serving as three legs. The inclination angle was determined at which the block
was sliding without acceleration on an acrylic flat surface (similar to the silo wall).
The friction coefficient was determined as the tangent of the inclination angle.
For clean glass beads and a clean surface, the friction coefficient was $\mu_\text{clean}=0.44$,
while for lubricated glass beads and a lubricated surface, we got a considerably reduced value 
$\mu_\text{lubricated}=0.015$.
The surface of the hydrogel beads is naturally wet, so for these grains the same measurement
was repeated without any lubricant, and resulted a friction coefficient of $\mu_p=0.02$.
We note that the typical normal force in these tests was much larger (a few hundred times)
than the weight of a single grain.
In the {\it second set} of experiments, a thin plastic foil was used instead of the wooden block,
thus the sliding object essentially had the weight of the three beads.
In this case the measured sliding friction coefficient of the clean glass beads was 
$\mu_\text{clean}=0.27$ for the smaller ($d=3.15$~mm), and $\mu_\text{clean}=0.18$ for the larger 
($d=7$~mm) beads.
For lubricated glass beads, stationary sliding was observed for a wide range of the plane inclinations
$35^o<\theta<60^o$ for $d=3.15$~mm, and $11^o<\theta<20^o$ for $d=7$~mm, corresponding to friction
coefficients of $0.7<\mu<1.7$ and $0.19<\mu<0.36$ for small and large beads, respectively.
The stationary sliding speed was typically a few cm/s, it increased with $\theta$.
Thus, for the case of glass beads for typical sliding velocities observed in our silo during 
discharge the friction coefficient at lubricated contacts changes considerably with the normal 
force. For certain contacts it can be similarly small to the case of soft hydrogel beads, for other 
contacts its value is much larger, it is similar to that of dry glass bead contacts.

\section{Numerical system}

Discrete Element Method (DEM) simulations were performed using LIGGGHTS \cite{klossPCFD2012}, a
general granular simulation software.
The DEM algorithm resolves the particle--particle interactions, and integrates both the 
translational and rotational motion of each particle. 
The interaction force ${\vec F}_{ij}$ between two contacting
particles $i$ and $j$ was computed using the Hertz-Mindlin model with no-slip condition, 
the default nonlinear implementation of LIGGGHTS \cite{direnzo2004comparison}.
This numerical scheme allows the estimation of the elastic and damping interaction parameters 
given the Young modulus $Y_m$, Poisson's ratio $\nu$ of the material, the normal 
restitution coefficient $e_n$ and friction
coefficient $\mu$ of the particles. Moreover, the magnitude of the elastic tangential force
$|{\vec F}_{ij}^{t}|$ is constrained by the  normal one, $|{\vec F}_{ij}^{n}|$, satisfying the
Coulomb constraint $|{\vec F}_{ij}^{t}|< \mu |{\vec F}_{ij}^{n}|$.
The effect of rolling resistance and resistance to torsion were not taken into account.

We use a cylindrical silo with similar dimensions as the experimental system,
spherical beads with a polydispersity of the grain diameter of $\pm10\%$ for
hard beads and monodisperse for the case of soft particles. 
Test runs indicated that for the case of hard grains, polydispersity was necessary, as for a 
monodisperse system increased velocity fluctuations were observed (most probably due to local spatial 
ordering). No such differences were observed for the case of soft grains. 
The discharge process was simulated for various values of the material parameters by
systematically changing the Young modulus, the normal restitution coefficient $e_{\rm n}$ and
friction coefficient  $\mu$, while a constant particle density $\rho_p = 1000\ \text{kg/m^3}$ 
and Poisson's ratio of $\nu = 0.45$ was used in all cases. 
For the case of hard grains we used $Y_m = 5 \cdot 10^8$ Pa which results in a negligible 
Hertzian normal deformation $\delta_n/d \sim (p_b/Y_m)^{2/3}= 4 \cdot 10^{-4}$ for a basal 
pressure of $p_b=4$ kPa. For soft grains, we used $Y_m = 1.25 \cdot 10^5$ Pa yielding 
$\delta_n/d \sim (p_b/Y_m)^{2/3}= 10^{-1}$. The particle-wall interaction was modeled using the 
same contact parameters used for particle-particle interaction.
The time steps were estimated to be less than $t_c/40$, where $t_c$ is the contact time.

For the preparation of the initial state of the simulations, the positions of the particles
were generated using the random sequential deposition model in order to avoid
overlaps between particles. For each configuration (interparticle friction and material
stiffness), the procedure was realized using two protocols: one with an interparticle
friction coefficient of $\mu_\text{F}=0.5$ and one with zero interparticle friction. 
This was done to mimic the fast and slow filling
procedures in the experiment, respectively. Here, filling the silo with zero interparticle
friction resulted in a slightly higher initial packing. After filling, the orifice is opened and
the material starts to flow out of the container.

Similarly to the experiments, we monitored two macroscopic quantities during the outflow 
process. First, the flow rate was calculated using the number of particles exiting the silo in 
fixed time intervals of $0.1$ s. Second, the vertical component of the total force acting on the
bottom surface was calculated. The mean height of the column was estimated averaging the highest 
particle locations within 8 equally spaced intervals in the radial direction. 

\subsection{Coarse-grained continuum description}

In order to extract the averaged macroscopic fields from the DEM data, a coarse-graining (CG) 
micro-macro formulation can be derived from the classical laws of conservation 
\cite{Goldhirsch2010,Weinhart2013,Richard2015,Artoni2019}.
Thus, this procedure leads to expressions for macroscopic density,  velocity  and stresses, in 
terms of particle properties and the contact interactions.

Starting with the DEM data for the positions and velocities of the particles,
the microscopic mass density of a granular flow $\rho(\vec{r},t)$ can be
defined~\cite{Goldhirsch2010,Weinhart2013,Richard2015,Artoni2019}
at time $t$ as

\begin{equation}
\rho\left(\vec{r},t\right) = \sum_{i=1}^{N} m_i \phi\left(\vec{r}-\vec{r}_i(t)\right),
\end{equation}
\noindent where the sum runs over all the particles within the system and
$\phi\left(\vec{r}-\vec{r}_i(t)\right)$ is an integrable coarse-graining function, which was
chosen to be a 3D Gaussian function with standard deviation equal to the mean particle 
radius $d/2$.
In the same way, the coarse-grained momentum density function $\vec{P}(\vec{r},t)$ is defined by
\begin{equation}
{\vec P}(\vec{r},t) = \sum_{i=1}^{N} m_i \vec{v}_{i} \phi\left(\vec{r}-\vec{r}_i(t)\right),
\end{equation}
where $\vec{v}_{i}$ represents the velocity of particle $i$.
The macroscopic velocity field $\vec{V}(\vec{r},t)$ is then defined as the ratio of
momentum and density fields,
\begin{equation}
\label{velocidad}
{\vec V}(\vec{r},t) = {\vec P}(\vec{r},t)/\rho(\vec{r},t).
\end{equation}

In order to define the mean stress field, we have used a very elegant and mathematically consistent
definition of mean stress ${\bar \sigma}_{\alpha \beta}$ introduced by Goldhirsch~\cite{Goldhirsch2010}.
Following this approach, the total stress field ${\sigma}_{\alpha \beta}$ is composed
by the kinetic stress  field $\sigma^k_{\alpha \beta}$  and the contact stress
field  $\sigma^c_{\alpha \beta}$. They are defined as follows. The mean contact
stress tensor reads as
\begin{eqnarray}
\sigma^c_{\alpha \beta}(\vec{r},t) =
 -\frac{1}{2}\sum_{i=1}^{N} \sum_{j=1}^{Nc_i} f_{ij \alpha} r_{ij \beta} \int_0^1  \phi(\vec{r} - \vec{r}_i + s \vec{r}_
{ij} ) ds ,
\label{contact_stress}
\end{eqnarray}
where the sum runs over all the contacting particles $i,j$, whose centers of mass are at
$\vec{r}_i$  and  $\vec{r}_j$, respectively. Moreover, $\vec{f}_{ij}$ accounts for  the force
exerted by particle $j$ on particle $i$ and $\vec{r}_{ij} \equiv \vec{r}_i - \vec{r}_j$.

The kinetic stress field reads as
\begin{eqnarray}
\sigma^k_{\alpha \beta}(\vec{r},t) = -\sum_i^{N} m_i v_{i \alpha}' v'_{i \beta} \phi\left(\vec{r}-\vec{r}_i(t)\right),
\label{kinetic_stress}
\end{eqnarray}
where $\vec{v}'_i$ is the fluctuation of the velocity of particle $i$, with respect to the mean 
velocity field.
\begin{equation}
\vec{v}_{i}' (t,\vec{r}) =\vec{v}_{i} (t) - \vec{V} (\vec{r},t).
\end{equation}

Once the total stress tensor $\sigma_{\alpha \beta}(\vec{r},t) = \sigma^k_{\alpha \beta}(\vec{r},t) +
\sigma^c_{\alpha \beta}(\vec{r},t)$ is found, the pressure at any location
$p(\vec{r},t)=\frac{1}{3} Tr(\sigma_{\alpha \beta}(\vec{r},t))$ can be calculated.

Based on the previous theoretical framework, we implemented a post-processing tool
which allowed us to examine all the micro-mechanical properties of the particulate flow.
Taking benefit of the cylindrical symmetry of the system, one can average quantities
within the azimuthal direction. As a result, the obtained macroscopic fields are
in cylindrical coordinates $r$ and $z$. We will denote averaged quantities as

\begin{equation}
X(r,z,t)=\displaystyle \frac{1}{2\pi} \int_0^{2\pi} X(\vec{r},t)d\varphi.
\end{equation}

\section{Results}

\subsection{Filling the silo}

\begin{figure}[ht!]
\begin{center}
\includegraphics[width=10.3cm]{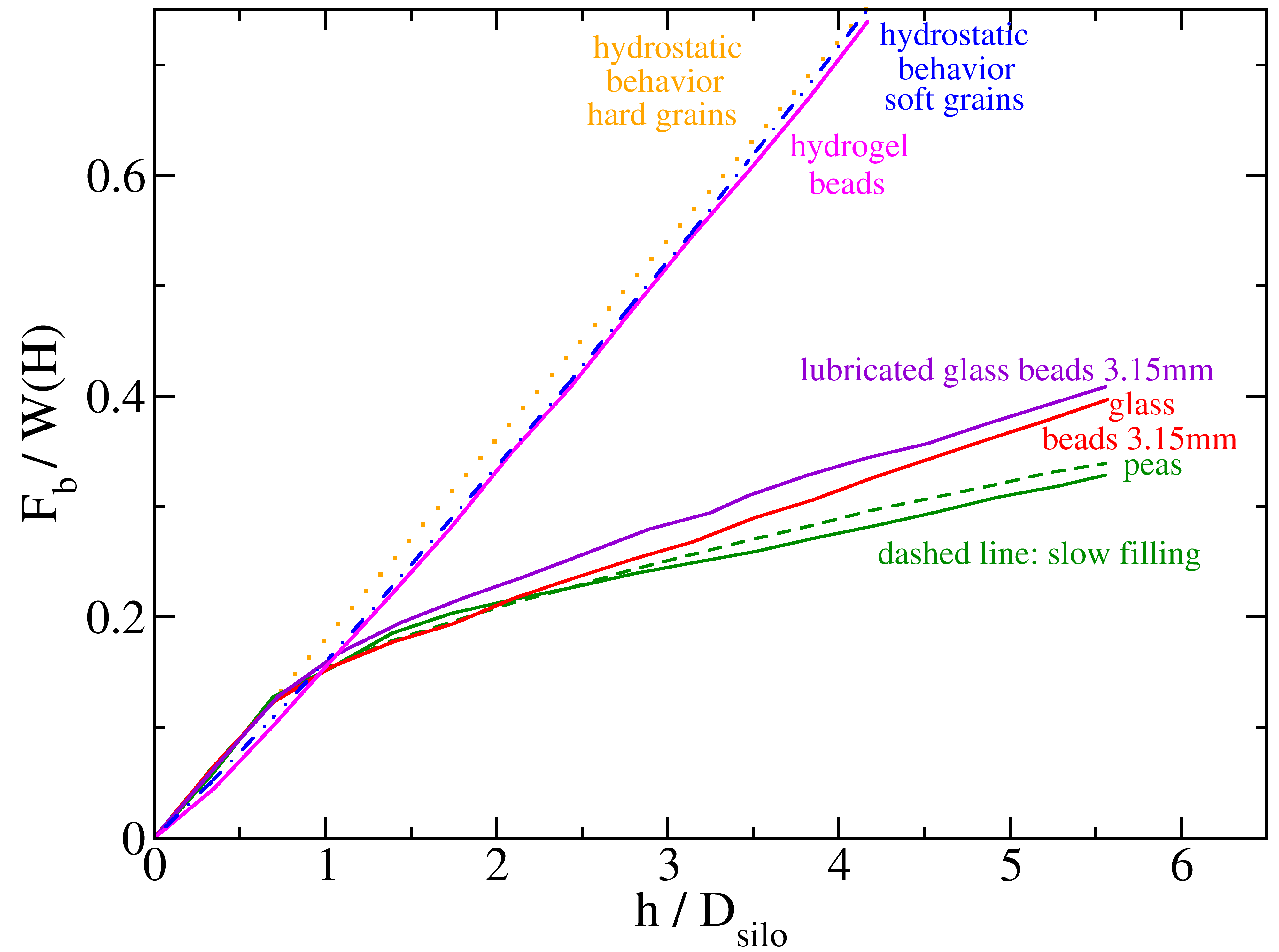}  
\end{center}
\vspace{-0.5cm}
\caption{
Experimental filling procedure: basal force $F_\text{b}$ as a function of the fill 
height $h$. For normalizing the basal force we use the total weight $W(H)$ of the sample 
which fills the silo entirely.
}
\label{janssen}
\end{figure}

As discussed in the introduction, it is long known that for a frictional granular material 
the force $F_\text{b}$ at the bottom of a silo deviates from the hydrostatic behaviour,
as part of the weight is supported by frictional contacts with the silo walls
\cite{janssen1895,sperl2006}. This effect is clearly seen for the frictional samples in 
Fig.~\ref{janssen}, where we normalize the basal force $F_\text{b}$ by the weight of the material 
$W(H)$ corresponding to a full silo. For peas, two curves are presented, corresponding
to the preparation protocols with filling speeds of 500 $\rm{cm}^3/\rm{s}$ (continuous lines)
and 50 $\rm{cm}^3/\rm{s}$ (dashed lines). The slower filling procedure (leading to denser
packing) resulted in slightly higher pressure at the bottom of the silo, i.e.~in this case
contact forces with the silo wall are weaker. In these measurements the silo was filled 
in 16 steps, and the filling height was measured visually in each step taking benefit of the 
transparent silo wall.

For hydrogel beads, the curve is very close to the hydrostatic behaviour, which
is coherent with the very low friction coefficient ($\mu_p<0.03$) of these particles.
Note that for this material the packing fraction increases with pressure, especially
at low pressures. This leads to the fact that for such a compressible material the
hydrostatic curve is not a straight line but its slope increases with $h/D_\text{silo}$, 
especially at filling heights $h\leqslant D_\text{silo}$.
For this material, we filled the container only up to a height of $60$ cm
($h/D_\text{silo}=4.17$), since at complete filling ($80$ cm) the pressure at the bottom of
the silo occasionally caused breaking of some beads, pushing fragments
into the gap between silo and bottom plate.

The $F_\text{b}$ curve for the lubricated glass beads is not similar to that of the hydrogel beads,
but almost the same as that of the dry glass beads.
This is due to the complex nature of lubrication, which is characterized by a contact 
force dependent friction coefficient as demonstrated in section \ref{exp}.
Moreover the thin layer of silicone oil also results in a cohesive interaction at the grain-grain
or grain-wall contacts. The magnitude of the cohesive force can reach the weight of a grain.
We have observed, that after discharge a few grains remain on the wall, and
slide down only very slowly.

\subsection{Silo discharge: comparison of experiments and simulations}

\begin{figure}[ht!]
\begin{center}
\includegraphics[width=\columnwidth]{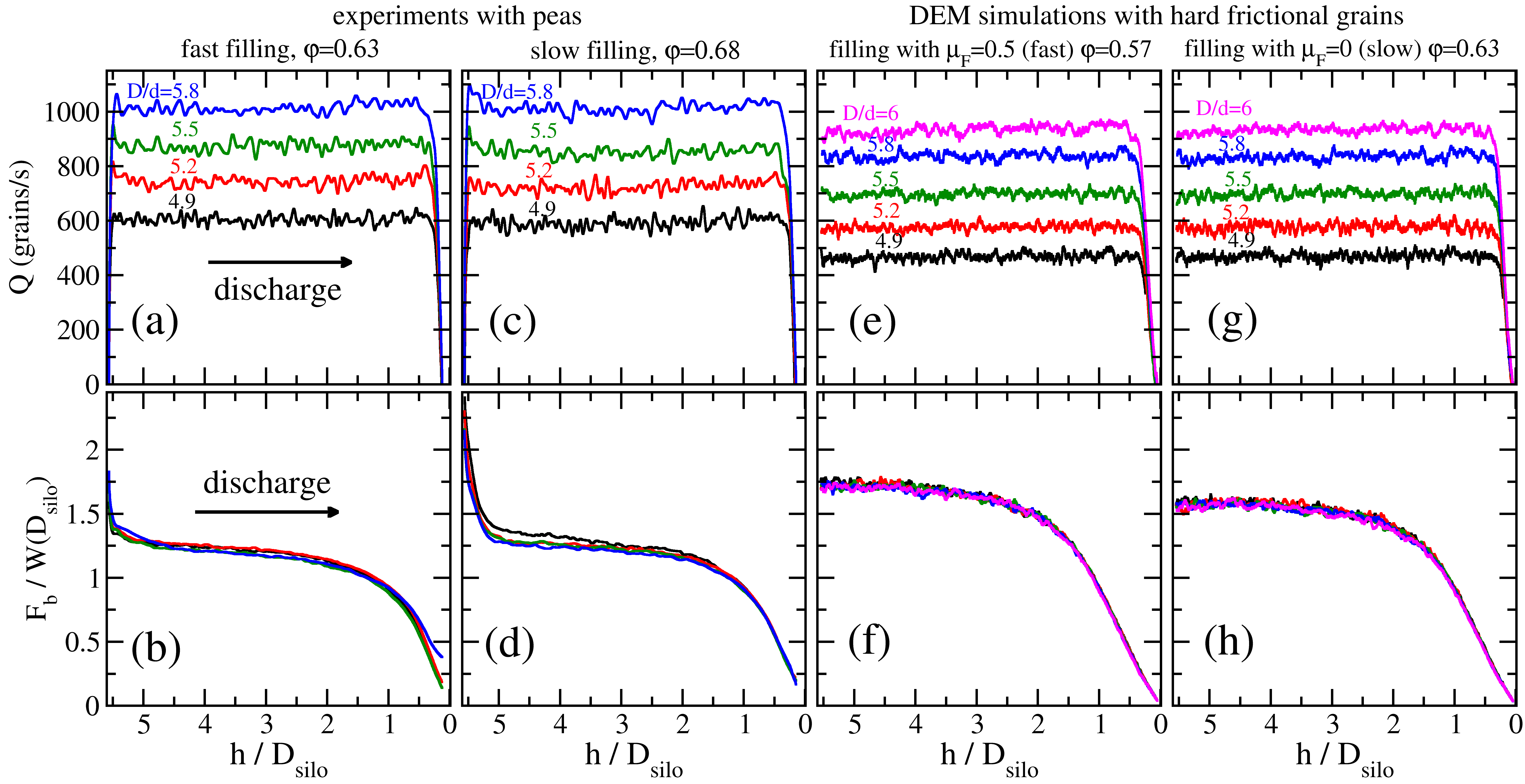}  
\end{center}
\caption{
Silo discharge of hard frictional grains: (a-d) experiments with peas and (e-h) DEM simulations
with spherical grains with friction coefficient of $\mu=0.3$, Young's modulus of 
$Y_m=5 \cdot 10^8$ Pa (inelastic) and a restitution coefficient of $e_\text{n}=0.9$.  
The top panels show the evolution of the flow rate $Q$, while the bottom panels present the  
normal force $F_\text{b}$ exerted on the silo bottom during the discharge process. 
The value of $F_\text{b}$ was normalized by the weight $W(D_\text{silo})$ of the material corresponding to a 
filling height of the diameter of the silo. The two types of initial conditions (slow and fast 
filling of the silo) result in very similar discharge processes both in experiment and simulation. 
The initial fill height was $h_0/D_\text{silo}=5.56$, while the initial packing fraction 
$\varphi$ is indicated at the top. 
Each curve represents the average of 4 measurements. The silo diameter was $D_\text{silo}/d=20.8$ 
(experiments) and $D_\text{silo}/d=15$ (simulations) with mean particle diameters of $d = 6.92$ mm in both.
}
\label{peas}
\end{figure}

The discharge curves, i.~e. the time evolution of the flow rate $Q$ and the force $F_\text{b}$
exerted at the bottom of the silo during discharge are presented for peas in Fig.~\ref{peas}(a-d) 
for both types of initial preparation (slow and fast filling). As is seen, both the flow
rate $Q$  and the basal force $F_\text{b}$ had a very similar time evolution for the two types
of preparation. Looking at the top panels, we see that the flow rate was basically
constant during discharge, even when the pressure conditions at the bottom of the silo
changed during the process.
The flow rate curves are the same for the two types of initial conditions,
i.e.~changing the initial density did not affect the discharge rate.

The bottom panels show that the basal force first decreased relatively quickly to
the same level for both initial conditions. This drop is partly attributed to the activation of the 
force network transferring weight to the container walls. After this initial transient -- which was
finished when about 5$\%$ of the total mass was discharged -- the basal force decreased only 
slightly until $h$ was approximately 2$D_\text{silo}$,
and then a faster decrease rate set in.
We note that the $F_\text{b}$ curves measured for different orifice diameters $D$ in 
Figs.~\ref{peas}(b) and (d) overlap, thus the basal force does not depend on the discharge rate.

We now compare the above experimental observations with the results of DEM simulations for hard 
beads with Young's modulus of $Y_m=5\cdot 10^8$ Pa and a restitution
coefficient of $e_\text{n}=0.9$. Similarly to the experiments, these results have been obtained
with frictional grains (friction coefficient: $\mu=0.3$) for both cases: filling with
$\mu_\text{F}=0.5$ and $\mu_\text{F}=0$, corresponding to fast and slow filling, respectively. 
As we see in Fig.~\ref{peas}(e-h), the numerical data reproduce all the features observed in the 
experiments: constant flow rate despite the changes in the basal force, same basal force 
curves for different orifice sizes (i.e.~for different discharge rates), and very similar values of 
the measured parameters for the two initial conditions. There is only a slight quantitative difference
between experiment and simulation: the flow rate is a bit lower and the basal force is
a bit higher in the simulations.

\begin{figure}[ht!]
\begin{center}
\includegraphics[width=12cm]{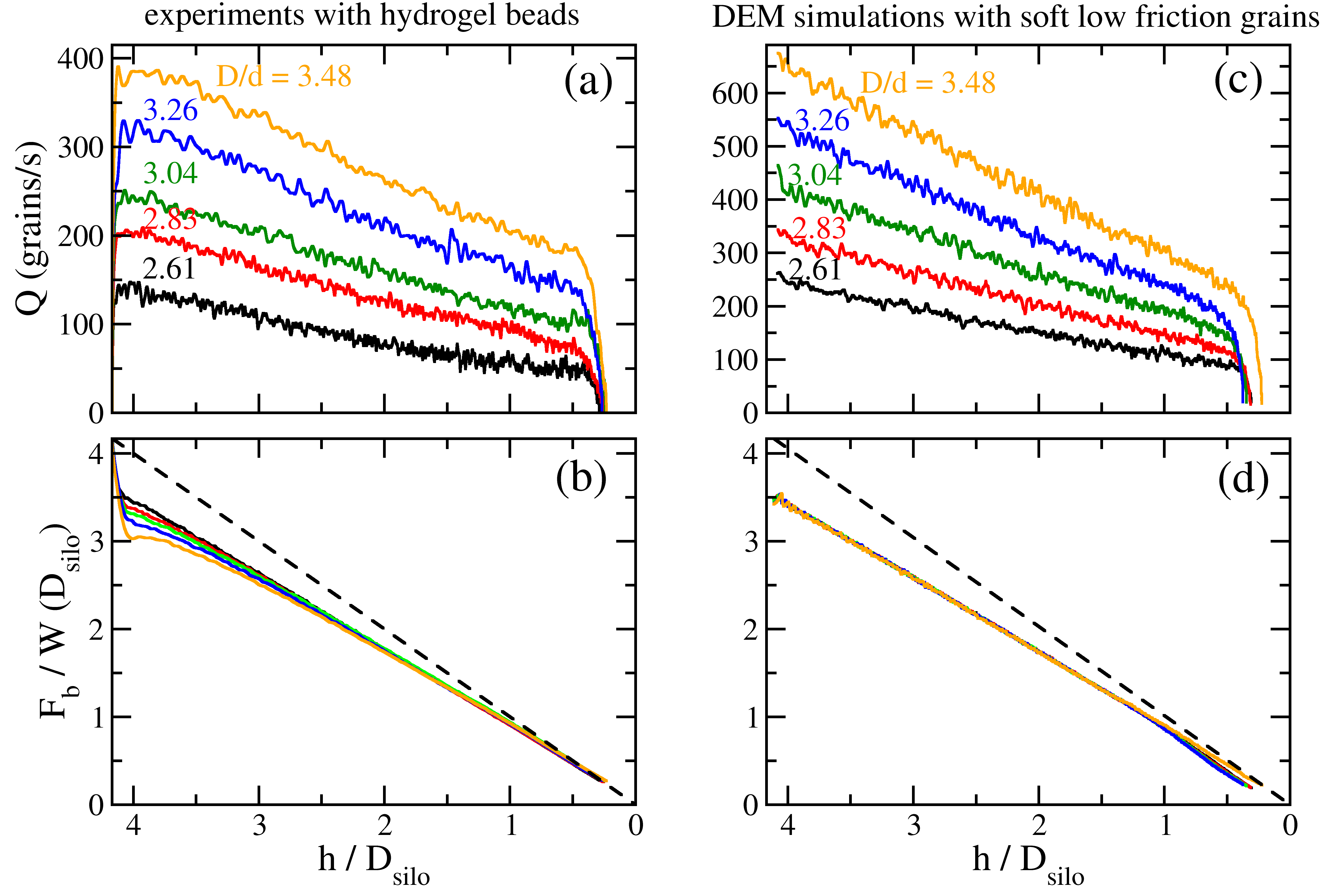}  
\end{center}
\caption{
Silo discharge of soft low-friction grains: (a-b) experiments with hydrogel beads and (c-d)
DEM simulations with spherical grains with friction coefficient of $\mu=0.03$, Young's modulus
of $Y_m=1.25 \cdot 10^5$ Pa and a restitution coefficient of $e_\text{n}=0.5$. The top panels show the
evolution of the flow rate $Q$ while the bottom panels present the basal force $F_\text{b}$ in 
the silo during the discharge process. The value of $F_\text{b}$ was normalized by the weight 
$W(D_\text{silo})$ of the 
material corresponding to a filling height of the diameter of the silo.  The dashed lines correspond
to the hydrostatic condition.
Experimental curves represent averages of 10 measurements, numerical data correspond to 
the average of 4 runs. Geometric parameters: $D_\text{silo}/d=15.65$,
$h_0/D_\text{silo}=4.17, d = 9.2\text{\ mm}$.
}
\label{hydrogel}
\end{figure}

The discharge of low friction, soft hydrogel beads is remarkably different from
the traditional granular discharge demonstrated above with peas and numerical
simulations for hard frictional grains. As seen in Figs.~\ref{hydrogel}(a-b),
the basal force $F_\text{b}$ decreases nearly linearly (close to the hydrostatic conditions),
and more importantly, the flow rate also decreases gradually during discharge. Another important 
thing to note is that for hydrogel beads clogging is only observed at much smaller orifice sizes, 
i.e. the silo discharge is continuous for the relatively small orifice sizes given in 
Figs.~\ref{hydrogel}(a-b), which is in accordance with earlier experimental observations 
\cite{stannarius2019-2}.  

In order to model the experimentally observed behaviour of
soft low friction grains, we performed a systematic numerical (DEM) study by changing the interparticle 
friction, the Young's modulus $Y_m$, and the coefficient of restitution of the particles.
By adjusting these parameters, the experimental data for hydrogel beads were best
reproduced with $\mu=0.03$,  $Y_m=1.25\cdot 10^5$ Pa and  $e_\text{n}=0.5$
(see Fig.~\ref{hydrogel}c-d). First, choosing $\mu=0.03$ ensures a quantitative match
between the experimental and numerical values of the basal force $F_\text{b}$ during discharge
(compare Figs.~\ref{hydrogel}b and d). Second, the numerical flow rate data
(Fig.~\ref{hydrogel}c) nicely reproduce the experimentally observed linearly decreasing
trend qualitatively, but are about $70\%$ higher than the experimental values
(Fig.~\ref{hydrogel}a). Increasing the Young's modulus and decreasing the restitution
coefficient both decrease the flow rate, with the Young's modulus having a stronger effect.
For this we use $Y_m=1.25\cdot 10^5$ Pa, which is about 3 times larger than the average nominal value.
This is in accordance with recent results by Brodu et al.~who showed, that when such soft
grains have multiple contacts, the resistance of the material against pressure becomes stronger,
as deformation of the grain due to the force at a given contact increases the contact force on
the other side of the grain \cite{broduPRE2016}. Better match between the flow rates of
experiment and simulation could not be obtained, since further increase of $Y_m$ (say by a
factor of 2) lead to clogging. 
The value of $Y_m = 1.25 \cdot 10^5$ Pa results in a typical Hertzian normal deformation of 
$\delta_n/d \sim (p_b/Y_m)^{2/3}= 10^{-1}$ for a silo with a basal pressure $p_b=4$ kPa.
As said above, changing the value of the restitution coefficient has a smaller effect on the flow rate, 
the value of $e_\text{n}=0.5$ was estimated from drop and bouncing experiments.

\begin{figure}[ht!]
\begin{center}
\includegraphics[width=\columnwidth]{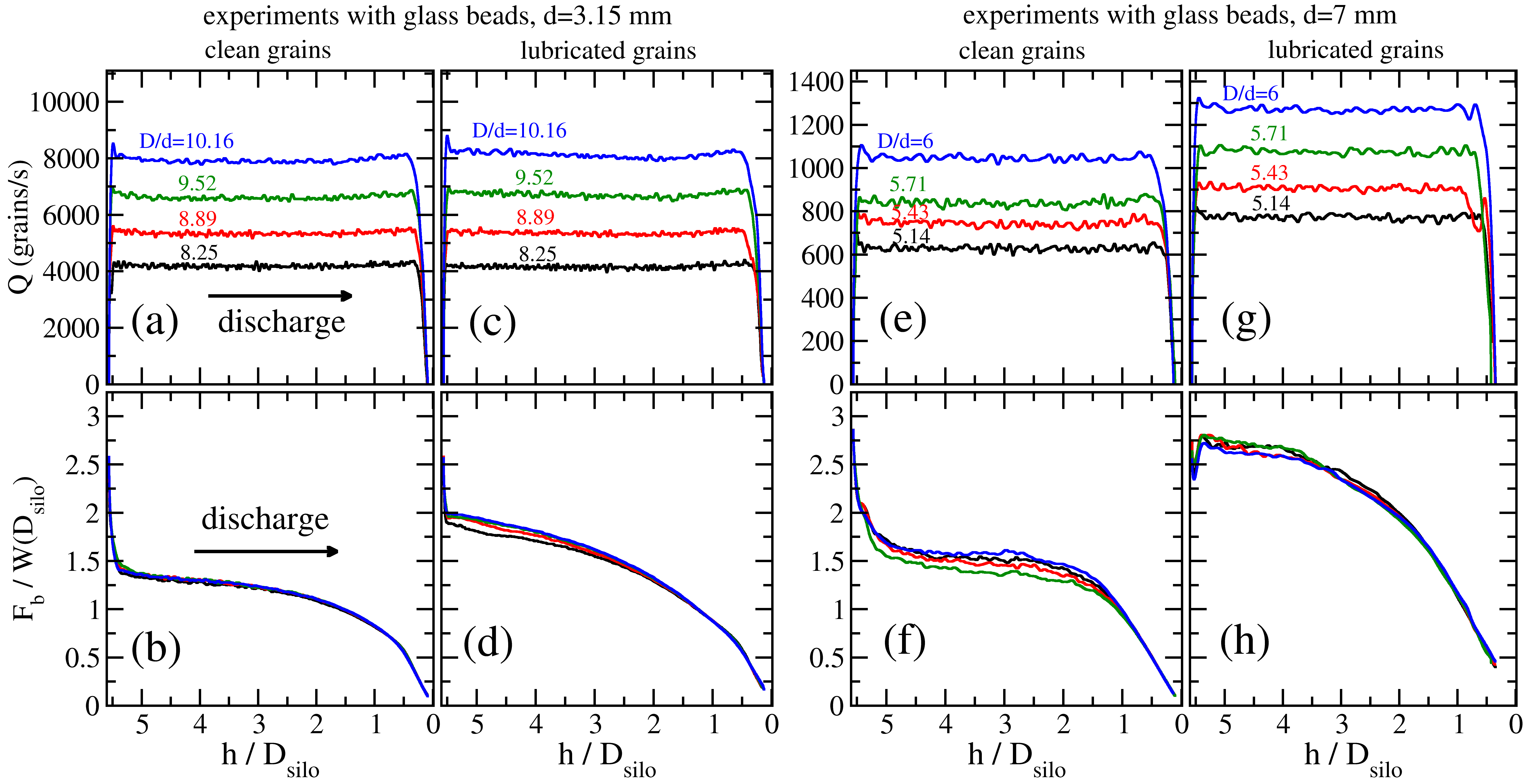}  
\end{center}
\vspace{-0.5cm}
\caption{
Silo discharge with clean and lubricated glass beads of (a-d) $d=3.15$ mm and (e-h) $d=7$ mm.
Top panels show the flow rate $Q$ while bottom panels correspond to the normalized basal force 
$F_\text{b}/W(D_\text{silo})$ as a function of height $h$ during the discharge process.
All curves represent the average of 4 measurements.
Other parameters: $h_0/D_\text{silo}=5.56$, 
$D_\text{silo}/d=45.7$ (for beads with $d=3.15$ mm) and
$D_\text{silo}/d=20.6$ (for beads with $d=7$ mm).
}
\label{glassbeads}
\end{figure}

In an attempt to test the effect of friction on hard grains’ discharge rates, we 
performed experiments with two more pairs of samples: clean (dry) and lubricated glass beads 
(using silicone oil) with diameters of $d=3.15$ mm and $d=7$ mm. As described in the experimental 
section, lubrication reduces the surface friction for certain contacts, and at the same time 
introduces new effects resulting from cohesive forces.
As expected, clean glass beads show very similar discharge characteristics
(see Figs.~\ref{glassbeads} (a-b) and (e-f)) to the above described peas.
For the lubricated systems, the results are slightly different for the smaller
($d=3.15$ mm) and larger ($d=7$ mm) grains. For the small grains, we observed the same 
flow rates for clean and lubricated systems (compare Figs.~\ref{glassbeads} (a) and (c)), 
only the basal force was a bit larger for the lubricated system. 
For the large beads, the discharge rate at a given orifice size is clearly
larger for the lubricated system (Fig.~\ref{glassbeads}(g)) than for the dry
system (Fig.~\ref{glassbeads}(e)), but it is still constant in time, like for
the small glass beads. Note that for the $7$ mm beads, lubrication lead to
a much stronger increase of the basal force than for the smaller ones.

\vspace{0.5cm}

\subsection{Effect of friction coefficient for hard and soft grains: DEM results and coarse-graining analysis}

\begin{figure}[ht!]
\begin{center}
\includegraphics[width=\columnwidth]{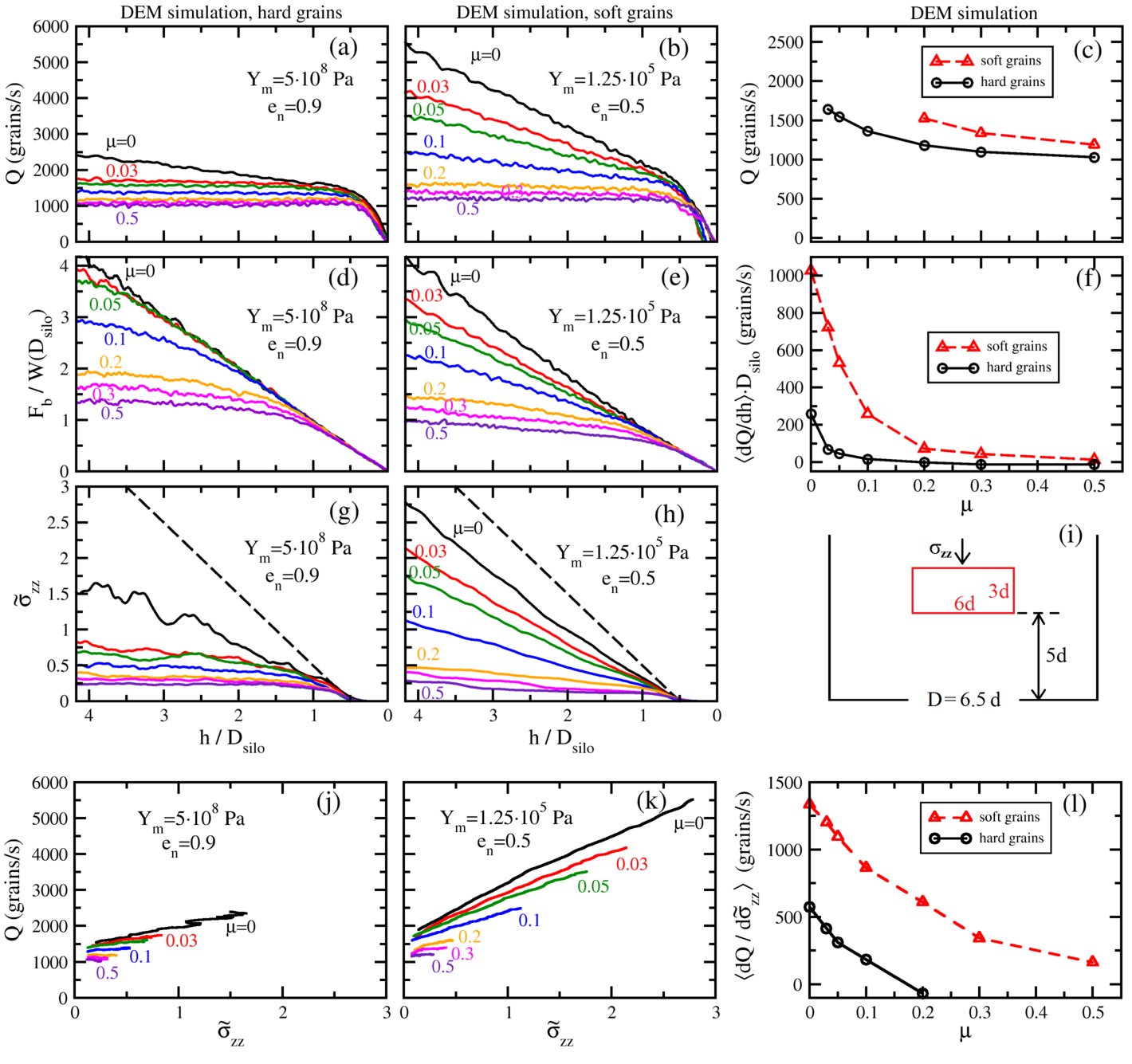}  
\end{center}
\caption{
DEM simulations of silo discharge for various values of the interparticle friction coefficient $\mu$.
Panels (a,d) show the evolution of the flow rate $Q$ and the normalized basal force 
$F_\text{b}/W(D_\text{silo})$ 
during discharge of hard particles with a Young modulus of $Y_m=5 \cdot 10^8$ Pa and a restitution 
coefficient of $e_\text{n}=0.9$ and (b,e) soft particles with $Y_m=1.25 \cdot 10^5$ Pa and 
$e_\text{n}=0.5$. Each curve corresponds to the average of 4 simulation runs.
Geometric parameters: $D/d=6.5$, $D_\text{silo}/d=15.65$, 
$h_0/D_\text{silo}=4.17$, $d = 9.2\text{\ mm}$.
(c) Initial constant flow rate and (f) the net gradient $\langle dQ/dh\rangle \cdot D_\text{silo}$ 
of the flow rate in the range of 
$D_\text{silo} < h < h_0$ from panels (a) and (b) as a function of the interparticle friction 
coefficient $\mu$.
(g-h) The normalized vertical stress $\tilde\sigma_{zz}$ measured in the region above the 
orifice marked with red in panel (i). The dashed line corresponds to the hydrostatic condition. 
(j-k) The flow rate $Q$ as a function of $\tilde\sigma_{zz}$.
Panel (l) presents the net gradient $\langle dQ/d\tilde\sigma_{zz} \rangle$ for both hard and soft 
grains.
}
\label{spheressimul1}
\end{figure}

Fig.~\ref{spheressimul1} demonstrates how the discharge behaviour changes when the interparticle 
friction is varied systematically. Two types of particles are examined: hard 
($Y_m=5 \cdot 10^8$ Pa, $e_n = 0.9$) and soft ($Y_m=1.25\cdot 10^5$ Pa, $e_n = 0.5$), while all other 
simulation parameters are kept constant. 
We see that decreasing friction coefficient leads
to a gradual change from the granular-like behaviour with constant flow rate
towards a behaviour characterized by a linearly decreasing flow rate.
The most important finding is, that this change is much stronger for soft grains
(Fig.~\ref{spheressimul1}b) than for hard grains (Fig.~\ref{spheressimul1}a), even though the
curves for $F_\text{b}$ do not differ that much for hard and soft grains (compare
Figs.~\ref{spheressimul1}d and e).
Thus, for soft grains decreasing interparticle friction leads to noticeable deviation
from the granular-like behaviour, while for hard particles such deviation is expected only
for frictionless grains. 
The height dependence of $F_\text{b}$ during discharge is similar for hard and soft grains, 
meaning that the vertical force transmitted to the wall does not depend significantly on the 
stiffness of the grains. The two experimental scenarios for the lubricated (low friction) glass 
beads with $d=3.15$ mm and  $d=7$ mm correspond to the hard grain simulation results with about 
$\mu=0.2$ and $\mu=0.12$, respectively.
For frictional beads (e.g.~$\mu=0.5$) the behaviour of hard and
soft grains is very similar with a slightly (1.17 times) larger flow rate for soft grains.
The effect of the interparticle friction coefficient on the flow rate is summarized in
Figs.~\ref{spheressimul1}(c,f). Panel (c) shows the average value of the flow rate $Q$ (for
those cases when it is constant during discharge), while panel (f) shows the net gradient 
$\langle dQ/dh \rangle \cdot D_\text{silo}$ 
of the flow rate (obtained by a linear fit in the range of $D_\text{silo} < h < h_0$)
as a function of $\mu$ for both hard and soft grains.

We can extract further data from the DEM simulations, e.g. more information about the 
stresses inside the silo. For this purpose, we use the coarse-graining methodology, which allows us
to build the macroscopic fields of the packing fraction
$\varphi\left(r,z,t \right) = \rho \left(r,z,t\right)/\rho_p$ (particle density $\rho_p$), 
velocity $\vec{V} (r,z,t)$ and stress tensor $\sigma(r,z,t)$, with all quantities averaged in 
the azimuthal direction. Since there is one-to-one correspondence between time and column height, 
similarly to  previous figures we will use the latter instead of the time parameter:
$X(r,z,t)\equiv X(r,z,h)$. In the following, we base our analysis on the average of four
simulation runs.

It is interesting to calculate the vertical stress $\sigma_{zz}$ above 
the orifice, averaged in a cylindrical region (marked with red in Fig.~\ref{spheressimul1}i). We 
normalize $\sigma_{zz}$ by the hydrostatic pressure $p_h = 4 W (D_\text{silo})/(D_\text{silo}^2 \pi)$ 
corresponding to the weight $W$ of a column of grains with a height of $h=D_\text{silo}$, similarly 
to the normalization of the basal force $F_\text{b}$. 
The evolution of this normalized vertical stress 
$\tilde\sigma_{zz}= \sigma_{zz} /  p_h$ is shown in Figs.~\ref{spheressimul1}g-h.
We find a stronger difference in $\tilde\sigma_{zz}$ between the case of hard and soft grains 
compared to the difference observed for the basal force. In fact, the $\tilde\sigma_{zz}$ curves 
change much less with $\mu$ for hard grains than for soft grains, just like the flow rate curves.
Thus, it is natural to look for correlation between the flow rate and $\tilde\sigma_{zz}$.
Figures ~\ref{spheressimul1}j-k present the flow rate $Q$ as a function of $\tilde\sigma_{zz}$.
While the data show a rather weak $Q(\tilde\sigma_{zz})$ dependence for hard grains, the increasing 
trend in $Q(\tilde\sigma_{zz})$ is clearly stronger for soft grains. The net gradient of the flow 
rate $\langle dQ/d\tilde\sigma_{zz} \rangle$ (obtained by a linear fit of the curves in 
Figs.~\ref{spheressimul1}j-k) is shown in Fig.~\ref{spheressimul1}l. Thus, for the case of soft 
grains, (i) the local vertical stress $\tilde\sigma_{zz}$ above the orifice changes more during the 
discharge process and (ii) the value of $\tilde\sigma_{zz}$ has a stronger impact on the outflow rate 
than for hard grains. Both facts (i) and (ii) contribute to the stronger $Q(h)$ 
dependence observed for soft grains. 

\begin{figure}[ht!]
\begin{center}
\includegraphics[width=\textwidth]{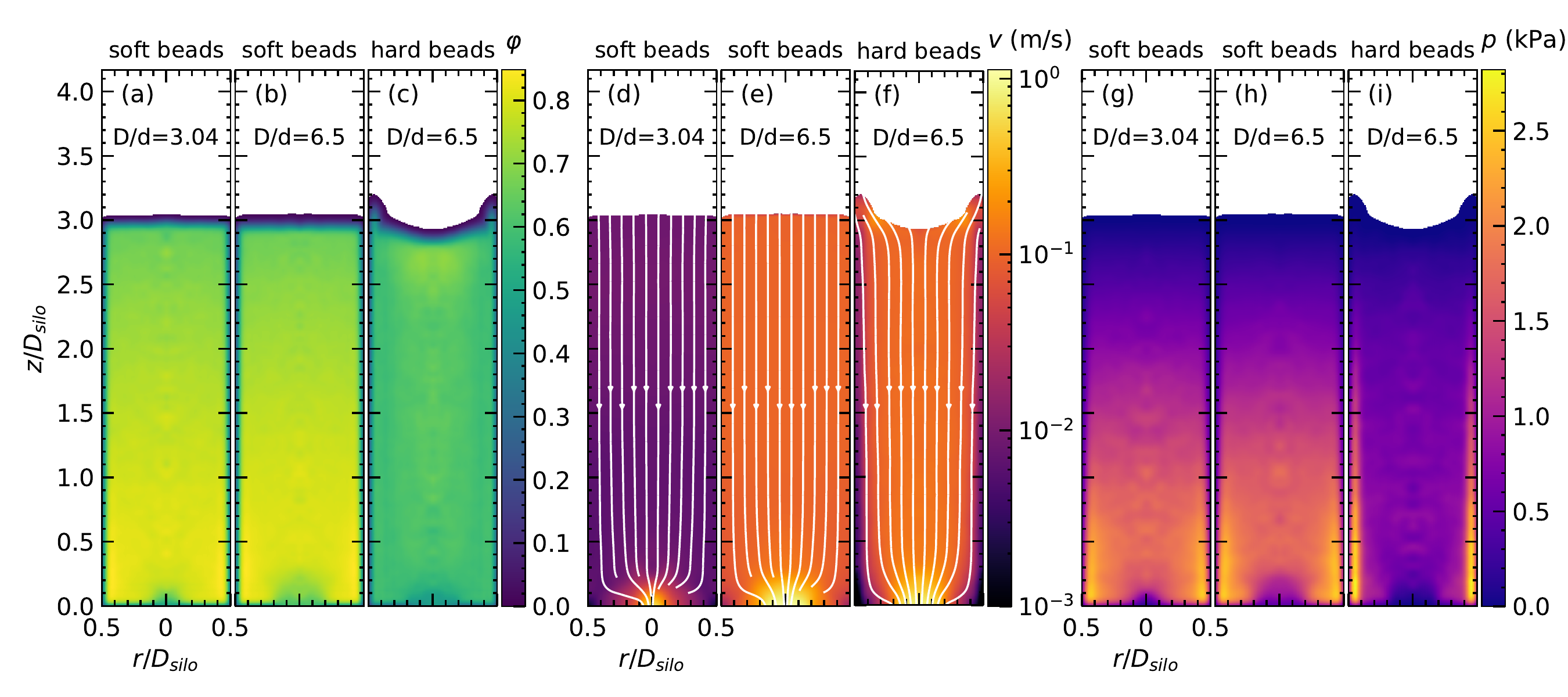}
\caption{Comparison of the discharge characteristics of hard and soft beads with friction 
coefficient of $\mu=0.03$: (a-c) color-maps of the volume fraction 
$\varphi\left(r,z,3 D_\mathrm{silo} \right)$, (d-f) the velocity field 
$v\left(r,z,3 D_\mathrm{silo} \right)$ and (g-i) spatial pressure profile
$p(r,z,3 D_\mathrm{silo})=\frac{1}{3} Tr(\sigma_{\alpha \beta}(r,z,3 D_\mathrm{silo}))$, 
obtained using the coarse-graining methodology.
Panels (a-b, d-e, g-h) correspond to the discharge of soft, while (c, f, i) to the discharge of hard 
particles with orifice sizes of $D/d = 3.04$ and $6.5$, as indicated on the plots.
Note that the state shown here for soft grains with small orifice size ($D/d = 3.04$) 
corresponds to the red curves in Fig.~\ref{vertical_profiles} and the color-maps for 
large orifices ($D/d = 6.5$) were obtained for the same 
parameters as the corresponding red curves of Fig.~\ref{spheressimul1}(a-b).
In the computation, we use a truncated Gaussian coarse-graining function $\phi(\vec{r})$ with
coarse-grained scale $w = d/2$. In all graphs, $z$ and $r$ are rescaled with the diameter of
the silo $D_\text{silo}$.
We use white color in all color-maps where the packing fraction is less than a cutoff value of 1\%. 
}
\label{color_maps}
\end{center}
\end{figure}

Figure \ref{color_maps} demonstrates the strong differences in the distributions of the volume 
fraction, velocity and pressure for soft and hard beads with a 
friction coefficient of $\mu=0.03$. For soft beads we present data for two orifice diameters $D/d=3.04$
and $D/d=6.5$ corresponding to the data presented in Fig.~\ref{hydrogel} and  Fig.~\ref{spheressimul1}, 
while for hard beads the data correspond to $D/d=6.5$.
Figures \ref{color_maps}(a-b) show color-maps exemplifying the volume fraction field
$\varphi\left(r,z,h \right)$ obtained in the simulations, for low friction soft particles
($\mu=0.03$,  $Y_m=1.25 \cdot 10^5$ Pa and $e_\text{n}=0.5$). Note, these parameters reproduce
the experimental flow rate data for hydrogel beads qualitatively (see Figs.~\ref{hydrogel} (c-d)).
The color-maps indicate that the volume fraction field $\varphi\left(r,z,h \right)$
is non-uniform, and in the top region the value of $\varphi$ resembles the random close
packing limit $\varphi \approx \varphi_\text{RCP}$. Examining lower regions, the system is more
compressed and $\varphi\left(r,z,h \right)$ increases until it reaches a maximum. Further down,
it decreases again upon approaching the orifice. On the other hand, Fig.~\ref{color_maps}(c) 
hows that the volume fraction field is much more homogeneous for the case of hard grains 
($Y_m=5 \cdot 10^8$ Pa and $e_\text{n}=0.9$).
Investigating the color-maps of the velocity field $v\left(r,z,h \right)$, for soft grains 
(Fig.~\ref{color_maps}(d-e)) we find that the velocity inside the silo is very uniform, with parallel 
streamlines. For hard grains (Fig.~\ref{color_maps}(f)) stronger inhomogeneities are observed in the 
velocity field with lower velocity near the walls and higher velocity in the central part of the silo,
leading to a surface distortion (dip) at the top of the granular column. We emphasize again, that the 
simulations for soft and hard beads were performed with the same friction coefficient $\mu=0.03$.
Finally, Figures \ref{color_maps}(g-i) display color-maps representing the spatial pressure profile 
$p(r,z,h)$,  obtained using the coarse-grained stress tensor, 
$p(r,z,h)=\frac{1}{3}Tr(\sigma_{\alpha \beta}(r,z,h))$. 
The cases for soft and hard grains are very different. The case of soft grains 
(Fig.~\ref{color_maps}(g-h)) is characterized by a gradual increase of the pressure with depth 
and relatively small variation in the radial direction. For hard grains (Fig.~\ref{color_maps}(i)) 
horizontal variations are very strong  and the change of pressure in the vertical direction is small 
in the central part of the silo. Thus, a large fraction of the weight of the granular column is 
supported by a few layers of grains near the wall.
This is the reason for the small variation of the vertical stress $\tilde\sigma_{zz}$ right above 
the orifice during the discharge process (Fig.~\ref{spheressimul1}(g)). We also note, that the three 
diagonal components (vertical, radial and tangential) of the stress tensor (not presented here) show larger
differences for hard grains, than for low friction soft particles and that our coarse-graining 
methodology has a limitation in accurately resolving the micro-macro transition near the walls.
Thus, for hard grains with a low friction coefficient ($\mu=0.03$), Janssen screening is still effective,
resulting in only a very slight change of the vertical stress above the orifice during the discharge
process.
This explains why the flow rate correlates with $\tilde{\sigma}_{zz}$ but not with
$F_b$ for the case of hard grains.

The above described observations are in accordance with previous experimental investigations 
which also revealed strong differences in the flow field between the case of plastic and hydrogel
beads using X-ray Computed Tomography \cite{stannarius2019,stannarius2019-2}. 
There the vertical velocity inside the 3D silo was found to be more homogeneous for low-friction 
soft hydrogel beads than for hard frictional plastic beads, and a decrease of the flow velocity 
was also observed with decreasing filling height using Ultrafast X-ray Computed Tomography 
\cite{stannarius2019-2}. 
Altogether, our current numerical data for the pressure distribution $p(r,z,h)$ for the case of 
soft grains with weak pressure change in the radial direction and a smooth pressure change along 
the symmetry axis of the silo allows us to perform some further analysis.
Namely, $p(r,z,h)$ gradually increases from the top of the bed along the symmetry axis, 
it has a maximum close to the orifice, and further down it decreases rapidly (see the color map at 
Figs.~\ref{color_maps}(g-h)), forming a measurable vertical pressure gradient right above the orifice. 
In the following, we will look for a correlation between the flow rate and this pressure gradient.

\subsection{Connection between the flow rate and pressure gradient near the outlet for soft, low-friction particles}
\begin{figure}[ht!]
\begin{center}
\includegraphics[width=12cm]{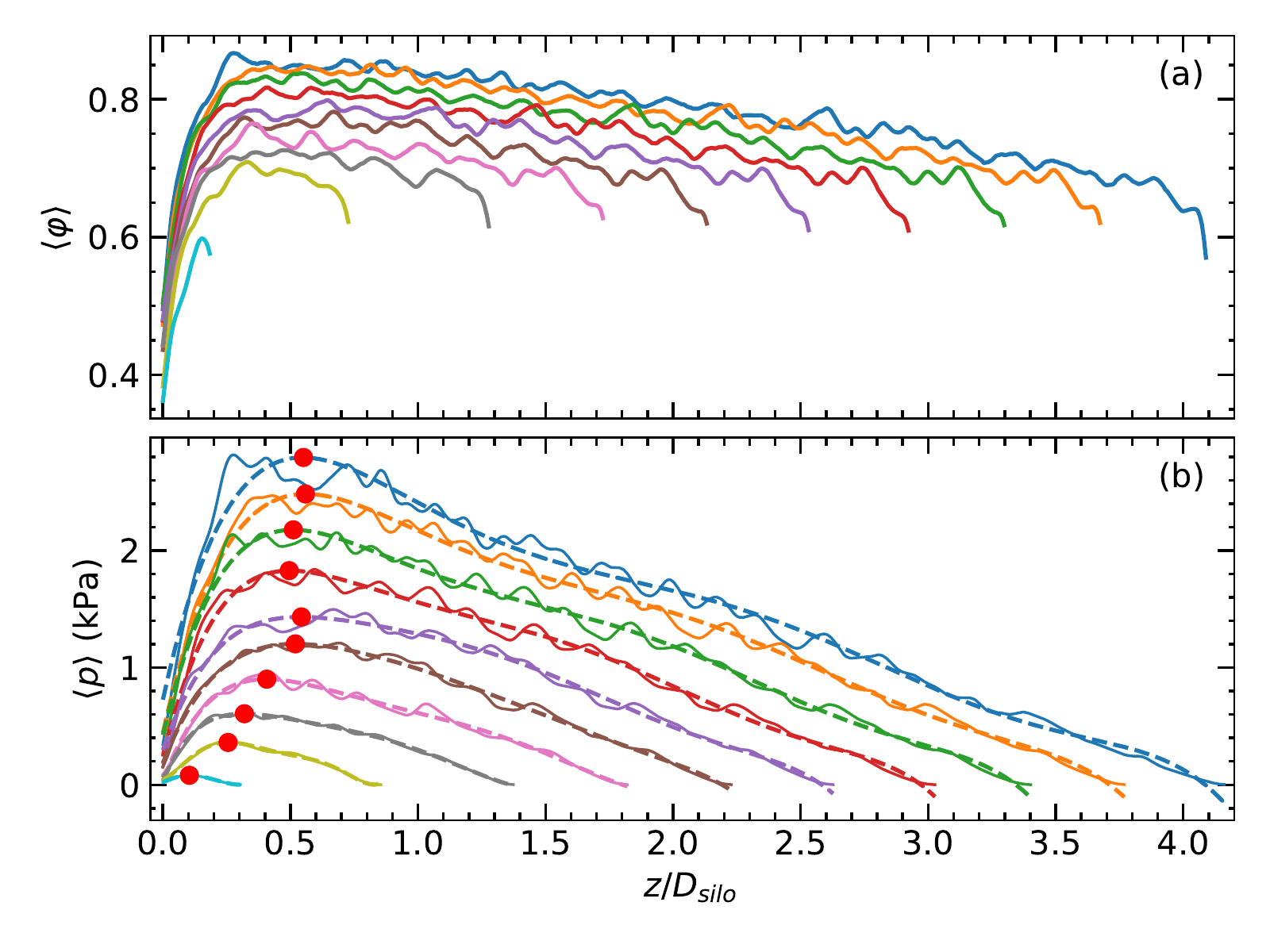}  
\caption{The profiles of (a) mean volume fraction $\langle \varphi \left(z,h\right) \rangle$ and (b) 
mean pressure $\langle p(z,h)\rangle$  obtained along the middle vertical slice of the 
silo. The curves correspond to successive instants of the discharge process. 
In both graphs, $z$ is rescaled with the diameter of the silo $D_\text{silo}$ and the relative size 
of the orifice is $D/d = 3.04$. In each case, the height of the column is given by the last point 
of the curve, approximately.
In panel (b), dashed curves are the 6th order polynomial fits for the corresponding mean 
pressure (solid), while the red dots mark the maxima of these fits.
}
\label{vertical_profiles}
\end{center}
\end{figure}

Aiming to explain the system's macroscopic response, in terms of its micro-mechanical properties,
Fig.~\ref{vertical_profiles} focuses the attention on the profiles of
$\varphi (z,h)\equiv\varphi \left(r{=}0,z,h\right)$ and $p(z,h)\equiv p(r{=}0,z,h)$ along the 
vertical direction, for soft grains with $\mu=0.03$ at the center of the silo with an orifice 
of $D=3.04d$. The data illustrated in Fig.~\ref{vertical_profiles} result from a
spatial average within a cylindrical region of one particle diameter in size, and they were obtained at
different instants in time, corresponding to subsequent stages of the discharge process.  
Figure \ref{vertical_profiles}(a) shows the profiles of the volume
fraction $\langle \varphi \left(z,h\right) \rangle$. 
In the top region, the volume fraction always reproduces the random close packing value.
These linear sections are simply shifted with respect to each other, which shows that the
density increases linearly with the depth, denoting the compression of the system. However,
density gradient remains practically invariant during the whole discharge process.
In all cases, the mass density shows a maximum close to the orifice, but its value drops
as time passes and the column gets shorter. Moreover, further down the dilatancy of the flow
becomes evident, the volume fraction is reduced, and at the orifice,
$\varphi(0,h)$ depends on the
column height $h$. 
A very similar trend appears for the mean pressure profiles
$\langle p(z,h)\rangle$ computed at different instants in time (see Fig.~\ref{vertical_profiles}(b)).
Starting from the column surface the pressure increases hydrostatically with depth, and
reaches a maximum value $p_{m}(h)$ at a certain vertical distance $L(h)$ from the orifice.
Note that the pressure in the center of the orifice $p(0, h)$ is nonzero, since the particles are still 
compressed at that point.

\begin{figure}[ht!]
\begin{center}
\includegraphics[width=13cm]{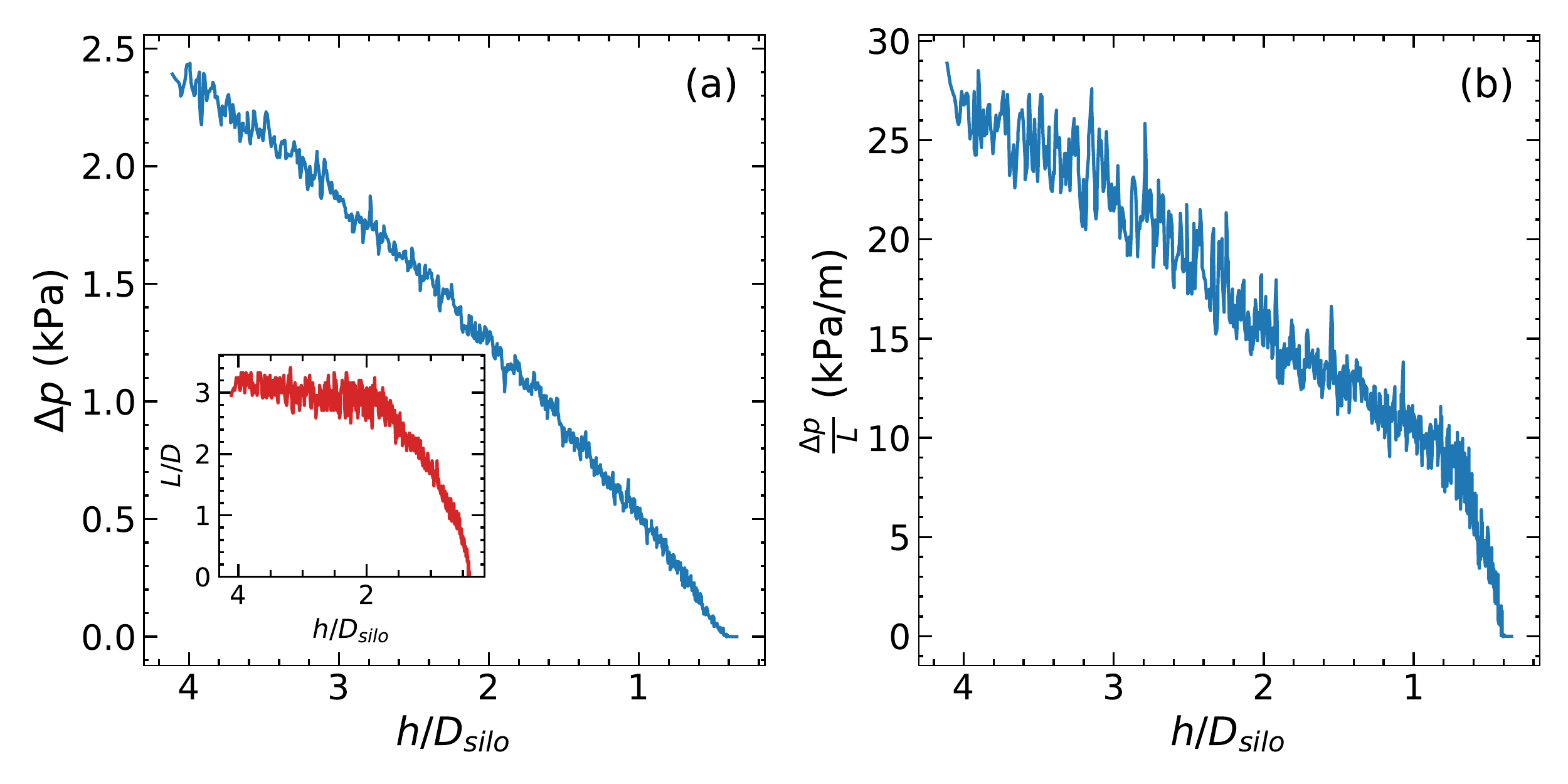}  
\caption{(a) Example of the pressure drop $\Delta p(h)  = p_m(h) - p(0, h)$ near the orifice as a function 
of the column height $h$. In the inset, the location of the maximum $L(h)$ respective to the 
orifice position is plotted as a function of the column height. In (b), the estimation of the 
pressure gradient $\frac{\Delta p}{L}(h)$ at the orifice as a function of the column height 
is shown. All data obtained for the orifice diameter $D/d=3.04$.
}
\label{fig:pressure_gradient}
\end{center}
\end{figure}

In order to address the forces acting on a representative volume element at the orifice, we
fitted 6th order polynomials to the pressure profiles $\langle p(z,h)\rangle$ along the 
vertical direction (Fig.~\ref{vertical_profiles}(b)). Then, we located the
distance of the maximum $L(h)$
from the orifice position, as well as the maximum pressure $p_m(h)$ (red dots) on these smoother curves,
as a function of the column height $h$ (shown in Fig.~\ref{fig:pressure_gradient}(a)). 
The pressure gradient at the orifice was estimated, using a linear approximation, by $\frac{\Delta p}{L}(h) = (p_m(h) - p(0, h))/L(h)$.
Two interesting things deserve to be commented here. First, during a large part of the discharge process, the
position $L(h)$ of the maximum pressure does not depend on the column height
(see inset of Fig.~\ref{fig:pressure_gradient}(a)). Accordingly, the pressure gradient
at the orifice (see Fig.~\ref{fig:pressure_gradient}(b)) decreases linearly.
Later on, however, finite size effects become more significant.

The coarse-grained continuous fields allow us to perform a theoretical analysis, to predict the
linearly decreasing flow rate
that was obtained numerically for low friction
soft particles and experimentally for hydrogel beads (see Fig.~\ref{hydrogel}).
Our arguments rest on the momentum balance that is established in steady state conditions at
the orifice, where the sum of the pressure gradient $\frac{\Delta p}{L}(h)$  
and the density of 
body gravitational force $\rho_p \varphi(0,h) g$ are balanced by the beads' resistance to motion.
Thus, assuming that the system acts like a viscous fluid in having a linear response to forcing,
the mean vertical velocity at the orifice can be predicted as 
\begin{equation}
v_z(D,h) =  K(D)  \left( \frac{\Delta p}{L}(h) + \rho_p \varphi(0,h) g_z\right),
\end{equation}
where $K(D)$ is a hydraulic coefficient that increases with decreasing effective viscosity of 
the system and increasing permeability of the orifice region.
Thus, the mass flow rate at a circular orifice of size $D$ reads
\begin{equation}
Q(D,h) =  \rho_p \varphi(0,h) A_c v_z(D,h) = \rho_p
\varphi(0,h) \pi \frac{D^2}{4} K(D)  \left(  \frac{\Delta p}{L}(h)  + \rho_p \varphi(0,h) g_z\right).
\label{flow_rate_theo}
\end{equation}

\begin{figure}[ht!]
\begin{center}
\includegraphics[width=10cm]{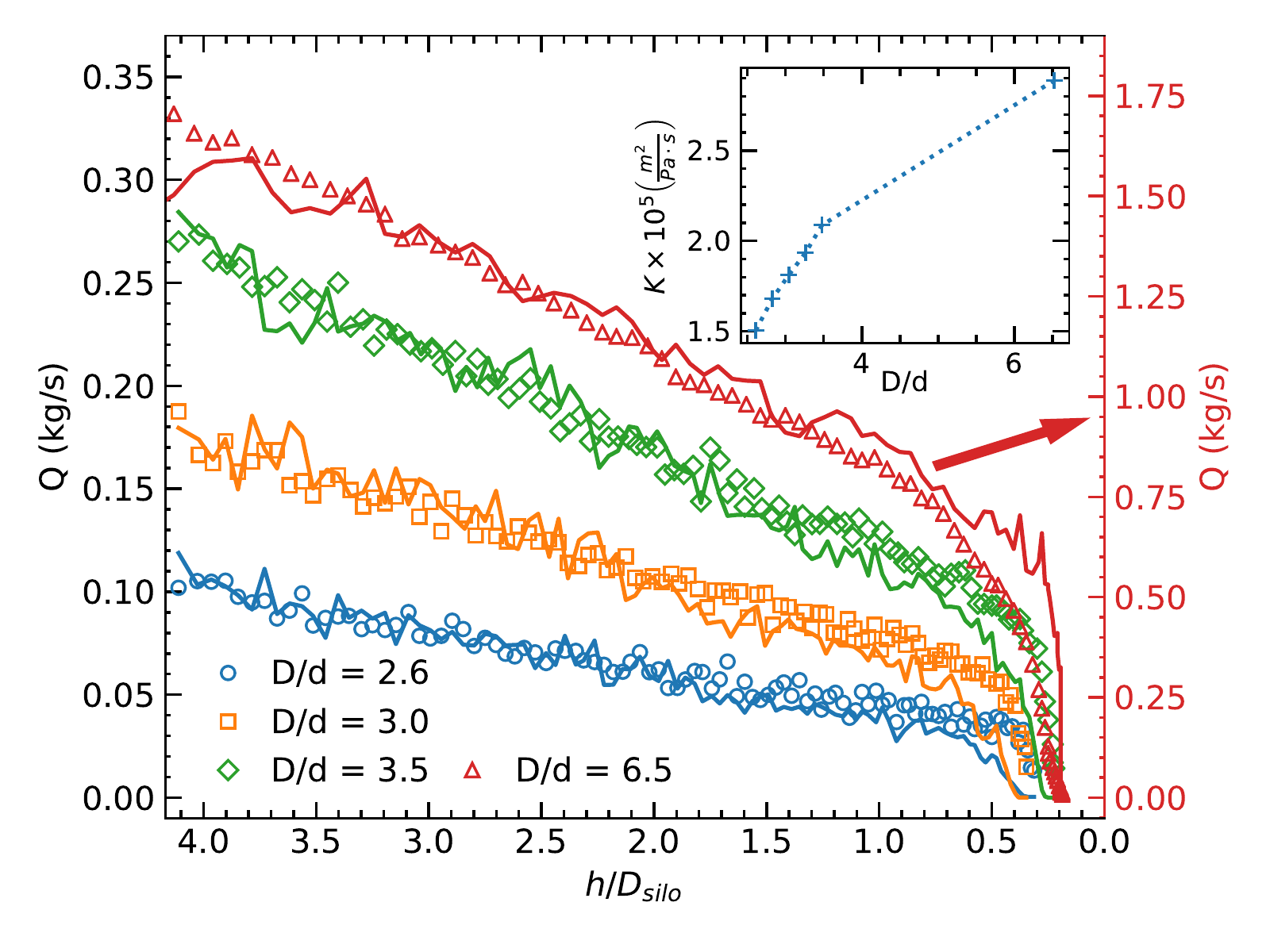}  
\caption{
Evolution of the flow rate $Q$ obtained numerically with particles of $\mu =0.03$,
$Y_m = 1.25 \cdot 10^5$~Pa, and $e_\text{n} = 0.5$ (symbols), for various values of the orifice 
diameter $D$. The bed height $h$ is rescaled with $D_\text{silo}$.
The numerical data are compared with the prediction of Eq.~(\ref{flow_rate_theo}) (lines),
where the pressure gradient $\frac{\Delta p}{L}(h)$ and the volume fraction $\varphi(0,h)$ at the 
orifice are derived from the coarse-grained fields.  Inset: the hydraulic coefficient $K$ as a 
function of orifice diameter $D/d$.  In the main figure, 
the $y$-axis labels at the right correspond to the data for
$D/d = 6.5$ (red), those at the left refer to the other three datasets.
}
\label{Q_compare}
\end{center}
\end{figure}

Figure \ref{Q_compare} compares the linearly decreasing trend of the flow rate with height,
obtained numerically for low-friction soft particles, with the theoretical prediction of
Eq.~(\ref{flow_rate_theo}).
Moreover, the inset of Fig.~\ref{Q_compare} shows the estimate of the hydraulic parameter
$K(D)$, which is the only fitting parameter used in the analysis. The agreement of the
comparison is remarkable, especially during the first part of the discharge process, where the
position $L(h)$ of the maximum pressure does not depend on the column height
(see inset of Fig.~\ref{fig:pressure_gradient}(a)) and the pressure gradient depends
linearly on  the column height.
Besides, we found that $K(D)$ increases with increasing orifice size.
Note, the value of $K(D)$ should be proportional to the permeability of the 
orifice and inversely proportional to the effective viscosity of the material.
The appearance of a linear response -- when the friction is low -- could be 
a natural consequence of the local interactions between the particles.
Namely, for deformable particles with low friction viscous normal damping has a more important 
role in energy dissipation than tangential frictional displacement, and
the macroscopic shear stress is proportional 
to the relative velocity of the neighboring particles, and so indirectly to the shear rate. 
Our analysis notably sheds light on the nature of the flow of a granular material consisting of
soft low-friction particles.
Indeed, its behavior is remarkably different from the 'traditional' granular discharge, where,
at the orifice, the pressure gradient is practically 
constant during the discharge process and, as a result, the volumetric flow rate does not depend on the 
column height \cite{rubiolargo2015}.

\section{Summary}
Our experimental and numerical investigations clearly show that changing particle stiffness has a 
strong effect on gravity driven granular flow out of a container with a small outlet in the flat 
bottom. Namely, decreasing the interparticle friction has much stronger effect for soft grains than 
for hard grains. For {\it soft grains}, with Young's modulus of the order of 10 times the basal 
pressure, numerical simulations predict that the character of the discharge process gradually 
changes with decreasing interparticle friction: Grains with high friction coefficient flow with a 
constant flow rate (granulate-like behaviour), while for grains with low surface friction, the 
flow rate systematically decreases with the height of the granular bed. This is noticeably different 
for {\it hard grains}, with Young's modulus of the order of $10^5$ times the basal pressure,  where 
basically a constant flow rate is observed except for the special limit case of frictionless grains. 
This is nicely demonstrated by the effect of lubricating the glass beads, where reducing 
the friction coefficient already resulted in a gradual decrease of the basal force during 
discharge, but the constant flow rate still persisted (traditional granulate-like behaviour).
For frictional grains, the difference between the discharge of hard and soft particles is 
smaller. Both cases are characterized with a constant flow rate, which is only slightly larger for 
soft grains. This is in accordance with earlier findings \cite{langston1995}, where for frictional
grains ($\mu=0.6$), a moderate change in the stiffness (typically 1 order of magnitude) lead to 
no significant change in the flow rate.
The evolution of the total normal force exerted at the bottom of the container 
during discharge shows much smaller differences between hard and soft grains.

Using DEM data combined with a coarse-graining methodology allowed us to compute
all the relevant macroscopic fields, namely, linear momentum, densities and stress tensors.
Such analysis reveals a considerably different pressure field for hard and soft grains 
with low friction coefficient. 
For low friction hard grains, Janssen screening is still observed with increased vertical stress 
near the walls, while for low friction soft grains this effect is much weaker.
For low friction soft particles the local vertical stress above the orifice changes more during 
discharge, and it has a stronger effect on the flow rate, than for low friction hard grains.
Thus, dynamic arch formation which is a key element in setting a constant flow rate for 
hard grains is much less effective for low friction soft grains. The fact that the critical orifice
size below which clogging is observed was found to be much smaller for low friction soft grains
\cite{ashourPRF2017,hong2017,harth2020} is in accordance with the above described observations.
For low friction soft grains, instead of the Janssen screening, the pressure inside the silo is 
linearly increasing with distance from the top surface and has a maximum right above the orifice.
The value of this pressure maximum gradually decreases during the discharge process, and can be 
related to the decreasing discharge rate.
Based on the momentum balance in the region of the orifice, we propose a phenomenological
formulation for soft particles with low friction coefficient. This suggests that
for such materials, dissipation is dominated by viscous friction near the orifice. 
This model predicts a very similar decrease of the flow rate with decreasing fill height as 
found in our experiments with hydrogel beads. 
We note, that incorporating a multiple-contact approach (such as in Ref. \cite{broduPRE2016}) 
to the DEM simulations is expected to further improve the quantitative match between experiment 
and simulation data, which might be the subject of future work.

\section*{Acknowledgements}
The authors acknowledge discussions with A. Ashour and V. Kenderesi.
Financial support by the European Union's Horizon 2020 Marie
Sk\l{}odowska-Curie grant ''CALIPER'' (No. 812638), by the NKFIH (Grant Nos.
OTKA K 116036 and 134199), by the DAAD/TKA researcher exchange program (Grant No.
274464), and by the BME IE-VIZ TKP2020 program is acknowledged.  
R.C. Hidalgo acknowledges the Ministerio de Econom\'ia y Competitividad (Spanish Government) 
Projects No. FIS2017-84631-P,  MINECO/AEI/FEDER,  UE. \\

\end{document}